# Bioinformatic analysis for structure and function of Glutamine synthetase（GS） from Antarctic sea ice bacterium Pseudoalteromonas sp. M175

**Jiahao Ma[1]\*, Guotong Xu[1], Le Ao[1], Siqi Song[1], Jingze Liu[2]**

[1]Harbin Institute of Technology, Weihai, China

[2]Qingdao University, Qingdao, China

2190790308@stu.hit.edu.cn

Abstract： Objective: To predict structure and function of Glutamine synthetase (GS) from Pseudoalteromonas sp. by bioinformatics technology, and to provide a theoretical basis for further study. Methods: Open reading frame (ORF) of GS sequence from Pseudoalteromonas sp. was obtained by ORF finder and was translated into amino acid residue. The structure domain was analyzed by Blast. By the method of analysis tools: Protparam, ProtScale, SignalP-4.0, TMHMM, SOPMA, SWISS-MODEL, NCBI SMART-BLAST and MAGA 7.0, the structure and function of the protein were predicted and analyzed. Results: The results showed that the sequence was GS with 468 amino acid residues, theoretical molecular weight was 51986.64 Da. The protein has the closest evolutionary status with Shewanella oneidensis. Then it had no signal peptide site and transmembrane domain. Secondary structure of GS contained 35.04% α-helix, 16.67% Extended chain，5.34% β-turn 、42.95% RandomCoil. Conclusions: This GU was a variety of biological functions of protein that may be used as a molecular samples of microbial nitrogen metabolism in extreme environments.

## 1. Introduction

Antarctic is a unique ecosystem on earth, which is composed of a combination of extreme cold, high salt, and strong radiations. Organisms of Antarctic have evolved their specialized cold and high salt tolerant enzymes to adapt to and survive in this harsh environment. In the work of others in this laboratory, a novel Glutamine synthetase strain Pseudoalteromonas sp. M175 (KU726544) was isolated and identified from Antarctic ice cover[1-2].

Considering the properties of cold-adaption protein at low temperature, low warm dependability and novel substrate specificities, along these lines, they have possible application in food manufacture, drug handling and basic life science. With the expanding worry of cold-adaption protein items, enormous scope maturation of the psychrophilic microorganisms is acquiring significance[3]. Synthetase by the traditional strategy includes transforming one autonomous variable keeping different elements steady, which include plenty of work. Nowadays, using a series of informatics tools can analysis the structure of the protein and identify whether it should be further studied or not, which is a effective choice.

In bacteria, glutamate and glutamine, which are mainly biosynthesized through ammonium assimilation, serve as the key nitro-gen donors for biosynthetic reactions. Glutamine is a significant ni-trogen source that is required for cell biosynthesis . Glutaminesynthetase is the most critical enzyme for nitrogen assimilation whennitrogen is limited; it is encoded byglnAinAgrobacteriumsp. Moreover,it utilizes ATP to convert glutamate and ammonia to glutamine . Itsactivity reduces as the intracellular nitrogen concentration increases[4].Due to the key role of glutamine synthetase in microbial nitrogen metabolism and the special ecological environment of Antarctica, it is necessary to carry out the bioinformatics analysis of glutamine synthetase of Antarctic sea ice bacteria[5].

The sequence of glutamine synthetase from sea ice bacterium Pseudoalteromonas sp., m175, obtained by laboratory PCR amplification and sequencing, was analyzed by bioinformatics databases and tools.The aim of these studies was to provide support for further theoretical research and application of microbial nitrogen metabolism under extreme environments.

## 2. Materials and methods

*2.1.sequence determination*

After 16s RNA gene sequence alignment learned that this strain was the closest relative to Pseudoalteromonas haloplanktis tac125, we used primerpremier5.0 software to design primers gs-f and gs-r based on the sequence of glutamine synthetase gene, and the synthesis and sequencing of PCR primers were done by SANGON biotech (Shanghai) Co., Ltd.
Primers:
GS-F：5'TAGAAACAGCCGTTATTACT3',
GS-R：5'CACCAGCGTAACTAGAAACAC3'
After that, the positive clones were obtained by PCR amplification of the target gene, preparation of the positive clones of the target gene, and other steps, and the positive clones were sent for sequencing by Shanghai SANGON biological Co., Ltd.

*2.2.Predicted methods*

Its open reading frame (ORF) and conserved domains were analyzed online using the ORF finder and conserved domains − search provided by NCBI; the physicochemical properties and Pro / hydrophobicity of glutamine synthetase from Antarctic sea ice bacteria were analyzed online using protparam tool and Protscale; the signalp 4.0 server tool was used to predict the signal peptide and localization signal, and tmhmm, respectivelyServer online predicted the transmembrane region of cohmgr1; the secondary structure of cohmgr1 was analyzed online with the sopma website, and its three-dimensional model was constructed with the SWISS-MODEL tool; the NJ phylogenetic tree was constructed with NCBI smart-blast with the Maga 7.0 software[4].

## 3. Results

*3.1. basic properties*
The results from ORF finder analysis at NCBI showed that this amino acid sequence had the longest ORF. Initiation codon was ATG, termination codon was TAA, and the full-length was 1408 bp (Figure 1). The full-lengh of amino acid sequence translated by DNA club was 468 amino acid residues (Figure 2) containing a complete conservative domain of Glutamine synthetase（GS）.

```
  1 M S Q S V L D F I K E N D V K F I D L R
  1 ATG TCG CAA TCG GTT TTA GAT TTT ATT AAA GAA AAT GAC GTT AAG TTC ATT GAT TTA CGC
 21 F T D T K G K E Q H I S I P H H Q I D E
 61 TTT ACT GAT ACA AAA GGT AAA GAG CAG CAT ATT TCA ATT CCT CAT CAC CAA ATT GAT GAA
 41 D F F E D G K M F D G S S I A G W K G I
121 GAC TTT TTT GAA GAT GGT AAA ATG TTC GAT GGT TCT TCA ATT GCT GGC TGG AAA GGC ATA
 61 N E S D M V L M P V A E S A K L D P F T
181 AAC GAA TCA GAC ATG GTG CTA ATG CCT GTT GCT GAA TCA GCT AAG CTT GAC CCA TTC ACT
 81 E E A T L I I R C D V V E P S T L Q G Y
241 GAA GAA GCC ACA TTA ATT ATA CGT TGT GAC GTA GTA GAG CCT TCT ACA TTA CAA GGT TAC
101 E R D P R S V A K R A E E Y M R S T G I
301 GAG CGC GAT CCA CGC TCT GTA GCA AAG CGT GCT GAA GAG TAT ATG CGC TCT ACA GGT ATT
121 A D T V L F G P E P E F F V F D D V K Y
361 GCT GAT ACG GTT TTA TTT GGC CCA GAG CCA GAG TTC TTC GTA TTT GAT GAC GTA AAA TAT
141 K T D M S G S M Y K I D S K Q A A W N S
421 AAA ACA GAC ATG TCT GGC TCT ATG TAC AAA ATC GAT TCT AAG CAA GCT GCT TGG AAC TCA
161 D K E Y A D G N T G H R P G V K G G Y F
481 GAT AAA GAA TAC GCA GAC GGC AAT ACA GGT CAT CGT CCT GGC GTT AAA GGC GGT TAC TTC
181 P V A P V D D F Q D W R S A T C L V L E
541 CCA GTT GCT CCT GTT GAT GAC TTT CAA GAT TGG CGC TCT GC T ACT TGT CTA GTA TTA GAA
201 E M G Q V V E A H H H E V A T A G Q N E
601 GAA ATG GGA CAA GTT GTT GAA GCA CAC CAT CAC GAA GTA GCA ACT GCA GGA CAA AAC GAA
221 I A T R F N T M V I K A D E I Q E M K Y
661 ATT GCT ACG CGC TTT AAT ACT ATG GTG ATT AAA GCC GAT GAA ATT CAA GAA ATG AAG TAT
241 V I H N M A H L Y G K T A T F M P K P I
721 GTT ATT CAT AAC ATG GCT CAT TTA TAC GGC AAA ACA GCC ACT TTT ATG CCT AAA CCT ATT
261 V G D N G S G M H C H Q S L A K D G V N
781 GTT GGC GAT AAC GGC TCT GGT ATG CAT TGT CAT CAG TCG TTA GCT AAA GAC GGT GTT AAC
281 L F A G D K Y G G L S E D A L Y Y I G G
841 TTA TTT GCA GGT GAT AAG TAC GGC GGT CTT TCT GAA GAT GCG CTT TAT TAC ATT GGC GGT
301 I I K H A K A I N A F A N A S T N S Y K
901 ATT ATT AAA CAT GCT AAA GCA ATT AAT GCG TTT GCT AAT GCC TCT ACT AAC TCG TAC AAA
321 R L V P G Y E A P V M L A Y S A R N R S
961 CGT TTA GTA CCA GGT TAC GAA GCA CCA GTA ATG CTT GCA TAC TCT GCA CGT AAC CGT TCT
341 A S I R I P V V P S A K G R R I E V R F
1021 GCA TCG ATT CGT ATT CCG GTA GTA CCA TCA GCT AAA GGT CGT CGT ATT GAA GTA CGC TTC
361 P D A T A N P Y L A F A A M L M A G L D
1081 CCT GAT GCA ACG GCT AAC CCG TAT CTT GCT TTT GCT GCC ATG CTT ATG GCT GGC TTG GAT
381 G I K N K I H P G D A M D K D L Y D L P
1141 GGA ATT AAA AAT AAA ATC CAT CCT GGT GAT GCA ATG GAT AAA GAT TTA TAC GAC CTA CCT
```

```
401 A E E A A E I P T V A S S L E E A L A S
1201 GCA GAA GAA GCA GCA GAA ATT CCA ACC GTT GCC TCT TCA TTA GAA GAA GCA CTG GCA TCT
421 L D A D R E F L N Q G D V F S N D L I D
1261 CTT GAT GCT GAC CGT GAG TTT TTA AAT CAA GGT GAC GTG TTC TCT AAT GAT TTA ATT GAT
441 A Y I K L K S Q E V E K L N M T T H P I
1321 GCT TAC ATT AAA CTT AAG AGT CAA GAA GTT GAA AAA CTT AAC ATG ACA ACG CAC CCA ATT
461 E  F  E  M  Y  Y  S  C  *
1381   GAG TTT GAA ATG TAT TAC AGC TGT TAA
```

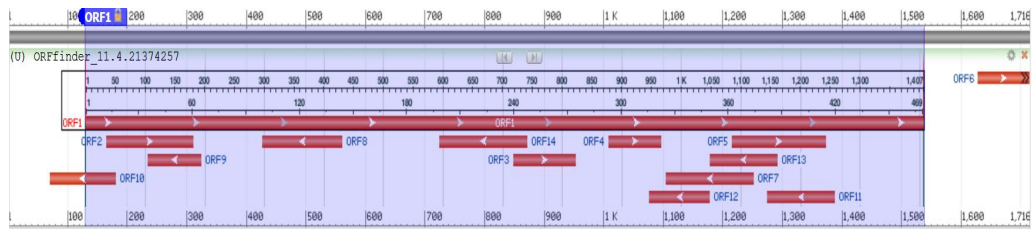

Fig. 1 NCBI output and sequence

The results were obtained using online prediction with ORF finder, online analysis with protparam tool and Protscale. The ORF finder results showed that this glutamine synthetase has a total of 468 amino acids, a molecular weight of 51986.64 Da, which isoelectric point is 4.93. The amino acid residue contents were shown in table1. Among them, negatively charged residues (ASP+Glu) total: 73 and positively charged residues (Arg + Lys) total: 47, belong to acidic proteins. This protein content can be detected spectrophotometrically by glutamine synthetase under polarized light at a wavelength of 280 nm, assuming that all Cys residues are cystine at a optical rotation of 45060 $M^{-1}cm^{-1}$ and assuming that all Cys residues are cysteine at a optical rotation of 44810 $M^{-1} cm^{-1}$.

Estimated half-life: the N-terminus of the considered sequence is m (MET). Estimated half lives are: 30 h (mammalian reticulocytes, in vitro); > 20 h (yeast, in vivo); 10 h (Escherichia coli, in vivo). Estimation of instability: Calculation of a destabilization index (II) of 46.39 allows to classify this glutamine synthetase as an unstable protein. Aliphatic index: 75.51, which was further based on the hphob / Kyte (5) Doolittle algorithm provided by the Protscale tool to analyze the Pro / hydrophobicity of the cohmgr1 protein. The results of the hydrophobicity analysis are shown in Figure 2. Overall, the hydrophobicity was low, indicating that the enzyme was more likely to be a water-soluble enzyme with a high average water (gravy): -0.344, which can be further informed that the enzyme was more stable.

Table 1 results of amino acid content analysis

| Amino acid classes | Number | Percent |
|---|---|---|
| Ala （A） | 47 | 10.0% |
| Arg （R） | 17 | 3.6% |
| Asn （N） | 18 | 3.8% |
| Asp （D） | 38 | 8.1% |
| Cys （C） | 4 | 0.9% |
| Gln （Q） | 12 | 2.6% |
| Glu （E） | 35 | 7.5% |
| Gly （G） | 32 | 6.8% |
| His （H） | 14 | 3.0% |
| Ile （I） | 29 | 6.2% |
| Leu （L） | 28 | 6.0% |
| Lys （K） | 30 | 6.4% |
| Met （M） | 19 | 4.1% |
| Phe （F） | 22 | 4.7% |
| Pro （P） | 22 | 4.7% |
| Ser （S） | 30 | 6.4% |
| Thr （T） | 20 | 4.3% |
| Trp （W） | 3 | 0.6% |
| Tyr （Y） | 19 | 4.1% |
| Val （V） | 29 | 6.2% |

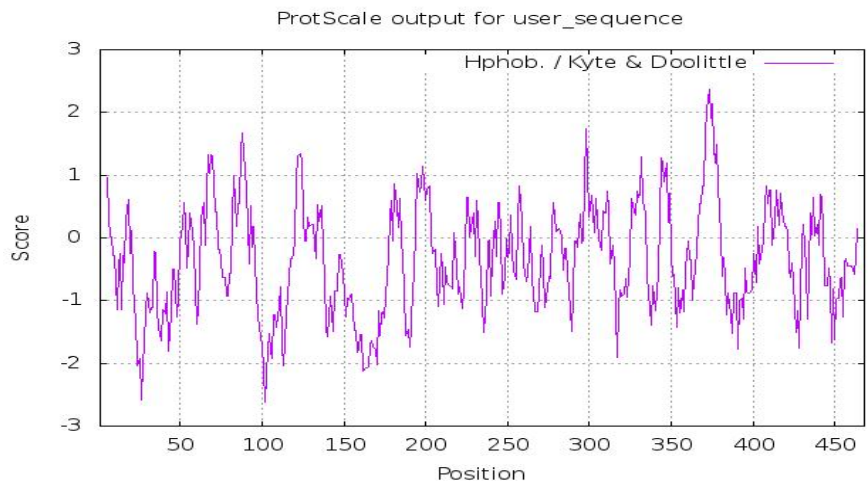

Fig. 2 hydrophobicity analysis

*3.3 Signal peptide, conserved domains and transmembrane regions*

The singalp 4.0 server and wolf PSORT were used to predict the secretion signal and localization of the protein online, respectively.Singalp 4.0 prediction results indicate that this protein has no signal peptide and is a non secreted protein.

The conserved domain of glutamine synthetase protein from Antarctic sea ice bacteria was analyzed online using CD-search provided by NCBI (http://www.ncbi.nlm.nih.gov/ncbi) and the results show that the encoded protein belongs to the glutamine synthetase superfamily and possesses a conserved sequence of glutamine synthetase from gram negative bacteria of about 470 amino acids, which confirmed our cloning results.The transmembrane regions were analyzed online using the tmhmm server too, this protein is a non transmembrane protein.

*3.4 Multiple sequence alignment and molecular evolution*

When the amino acid sequence of Antarctic sea ice bacterial glutamine synthetase was subjected to blsatp to search for similar sequences, the amino acid sequence of Antarctic sea ice bacterial glutamine synthetase was compared with GS of Shewanella oneidensis (genebank no.wp_011074057.1) with up to 99% sequence similarity.Prokaryotic glutamine synthetase amino acid sequences that had been reported in the NCBI database were selected and the phylogenetic (NJ) tree was predicted based on the neighbor joining method using NCBI SMART-BLAST. Results as shown in Fig. 9, Antarctic sea ice bacteria had high homology of GS proteins with the genera Shewanella and Neisseria, while they had low homology with those of Synechocystis.

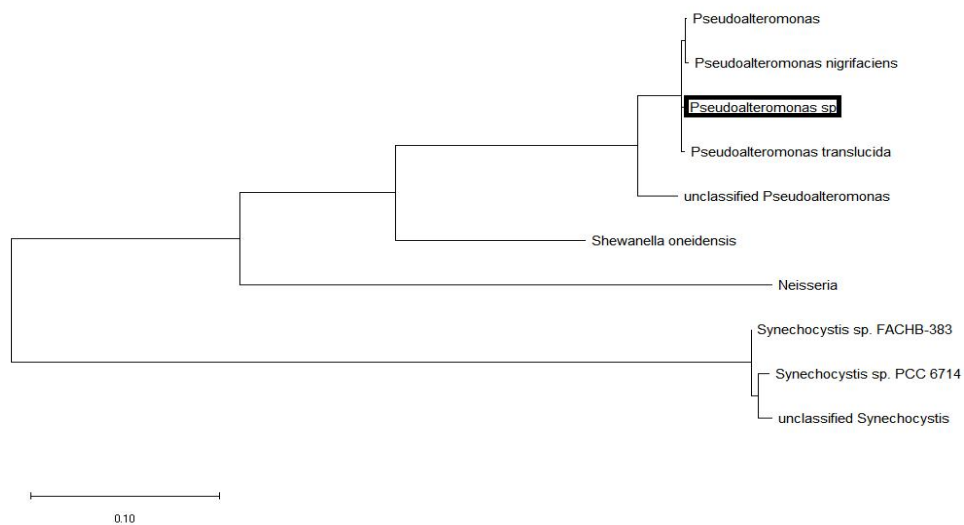

Fig. 3 molecular evolution tree construction results

*3.5 Secondary structure and tertiary structure*

The secondary structure of the protein was analyzed using sopma online software, which revealed that the secondary structure of glutamine synthetase contained 35.04% α - helix (α - helix, 72aa), 16.67% extended strand (extended, 92aa), 5.34% β - turn (β - turn, 33aa), 42.95% random coil (randomcoil, 188aa), indicating that the secondary structure of this glutamine synthetase is mainly random coil.

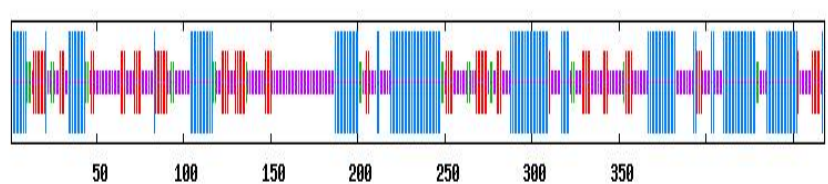

Fig. 4 secondary structure prediction of glutamate synthase

Notes: Blue: α - helix (alpha helix) green: β - Sheet (beta turn) yellow: random coil Red: extended strand

In the SWISS-MODEL modeling results, there were 126 templates that matched the target sequence.Upon screening, 29 were selected for modeling.The first five of them with a high match are shown in Table 4, and the modeling results are shown in Figure 8, exhibiting a regular hexagonal structure with holes in the middle, which are speculated to be the active sites.On prediction accuracy, gmqe (global estimate of model quality) was 0.89; qmean was -1.97, and overall the degree of match to experiment is likely to be high.

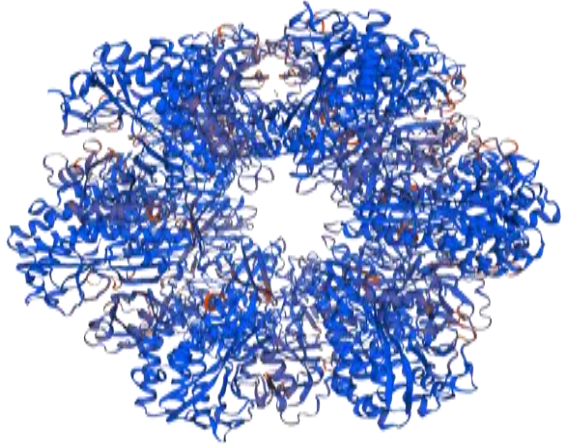

Fig. 8 SWISS-MODEL modeling results

## 4. Conclusion

GS are found universally in fungi, animals and plants[6], and have a function in the transmission of glutamate to glutamine，which is a significant progress of Extreme environmental adaptation. This work is a ahead research to identify the possible structure, function and the principle of molecular cold-adaption, and the evolution relation is partly declare in the Fig. 3.

But only the bio-informatics method and software is not enough to discover the whole progress of this problem. So, the chemical analysis and the spectrum is necessary[7]. Additionally, it is meaningful to analysis more Antarctic sea ice bacteria and more protein type have a whole view of marine microbe cold-adaption mechanism.

To sum up, the SWISS-MODEL modeling results and the sequence alignment is most valuable part of this work and it can be a reference for further research and it still have many shorten need to be completed.

## Conflict of interest statement

We declare that we have no conflict of interest.

## Acknowledgments

This work was supported by Special funds from the major one year project plan of Harbin Institute of Technology.

The teachers and classmates from the College of marine science and technology, Harbin Institute of Technology (Weihai), who provided help for the study, are gratefully acknowledged.Special thanks go to teacher Xioafei Wang for careful guidance and assistance with the project.